\documentclass[
a4paper,
onecolumn,
12pt,
titlepage,
nofootinbib,
amsmath,
amssymb,
floatfix,
showkeys,
]{revtex4-2}

\usepackage{graphicx}
\usepackage{multirow}

\usepackage{setspace}
\usepackage{color} 		
\definecolor{blue}{rgb}{0.0,0.0,1.0} 		
\definecolor{blue-green}{rgb}{0.0,0.5,0.5} 		
\usepackage[colorlinks]{hyperref}
\hypersetup{pdftitle={Energy Conditions in a Modified Brans-Dicke Theory},
	pdfauthor={Hootan Amani and Paul Halpern},
	linkcolor=blue-green, urlcolor=blue, citecolor=blue, filecolor=black, linktoc=page}
\begin{document}

\title{Energy Conditions in a Modified Brans-Dicke Theory}

\author{Hootan Amani}
	\email{hootan.amani@gmail.com}
	\affiliation{Department of Physics, Shahid Beheshti University, G.C., Evin, Tehran 19839, Iran}

\author{Paul Halpern}
	\email{p.halper@usciences.edu}
	\affiliation{Department of Mathematics, Physics, and Statistics, University of the Sciences in Philadelphia,\\ 600 South 43rd Street, Philadelphia, Pennsylvania 19104, USA}

\received{November 19, 2021}
\accepted{June 22, 2022}
\published{June 28, 2022}

\begin{abstract}
\noindent We consider a modified version of Brans-Dicke theory (MBDT) in four dimensions (4D) obtained by applying the induced matter method of Wesson to a 5D generalized Brans-Dicke theory. In 5D the model consists of pure vacuum, with no self-interacting potential, except for a scalar field. Following Wesson’s protocol, we group geometric terms in the 5D Einstein tensor arising from the extra dimension, move them to the other side of the generalized field equations, and identify them as the energy-momentum of the induced matter in 4D. Thus the extra dimension in 5D leads naturally to an effective matter field in 4D. Constraining the 5D geometry to be a generalization of the anisotropic Bianchi type I universe model first studied by Kasner, we derive the induced energy-momentum in MBDT and apply it to the investigation of energy conditions. The specified induced energy-momentum of that MBDT model consists of the energy density and directional pressure which indicate the anisotropy of the universe. We discuss the energy conditions and their bounds in the MBDT with such an induced imperfect fluid, with an eye toward a realistic model of the present-day universe, and consider the large-scale behavior of that spatially homogeneous and anisotropic model. We discuss how the energy conditions would be satisfied or violated in the context of MBDT, with the aim of providing a feasible description of the universe in the current era.
\end{abstract}


\maketitle

\tableofcontents
\newpage

\section{\label{sec:1}Introduction}
Although Einstein’s general theory of relativity is a masterful and accurate description of gravitation, he did not achieve all of his theoretical goals. Consequently, he and others have attempted variations of the original theory throughout the decades. One of Einstein’s goals in formulating the theory was trying to incorporate Mach’s Principle, in which inertia arises naturally from the large-scale material configuration of the universe. In obtaining cosmological solutions, he came to realize that standard general relativity does not achieve that aim. Consequently, in 1961, Carl Brans and Robert Dicke formulated a modified version of the theory, called Brans-Dicke (BD) gravity in which the gravitational constant is replaced with a changing scalar field \cite{Brans1961}. Note that neither the original general theory of relativity, nor BD theory, explain the present-day acceleration of the universe, unless a cosmological constant term, or a similarly acting source term, is included within the field equations.

Another of Einstein’s unrealized goals was to unify electromagnetism with gravitation. In that domain, he was inspired in part by Kaluza’s 1921 proposal of extending general relativity by a fifth dimension, and then incorporating a ``cylindrical condition'' (a set of specific mathematical constraints) to render the fifth dimension not directly observable \cite{Kaluza1921}. While Kaluza’s model (and later attempts by Einstein and others) lacked realistic solutions, it paved the way for modern higher-dimensional unification attempts.

Aside from the artificial ``cylindrical condition'', there have been numerous other attempts to explain the non-observability of any fifth dimension. One such model, based on a 5D extension of the Bianchi type I anisotropic cosmology studied by Kasner, was proposed by Chodos and Detweiler in 1980 \cite{Chodos1980}. In that model, anisotropic dynamics explains the disparity between the extra scale factor, and the other three spatial scale factors.

Yet another of Einstein’s philosophical critiques of the general relativistic field equations is that material and energetic source terms are needed on one side (in the form of the stress-energy tensor) to support the dynamics, along with pure geometric terms on the other side (in the form of the Einstein tensor). As he wrote in 1936:
\begin{quotation}
	``[GR] is sufficient --- as far as we know --- for the representation of the observed facts of celestial mechanics. But it is similar to a building, one wing of which is made of fine marble (left part of the equation), but the other wing of which is built of low-grade wood (right side of the equation). The phenomenological representation of matter is, in fact, only a crude substitute for a representation which would do justice to all known properties of matter~\cite{Einstein1936}.''
\end{quotation}

That discrepancy between the ``fine marble'' of geometry and the ``low-grade wood'' of matter inspired Wesson to propose the ``induced matter'' idea, which suggests that the additional geometrical terms arising from the extra dimension in the 5D Einstein equations can be associated with an effective energy-momentum tensor in 4D theory. In other words, the dynamics of the observed 4D non-vacuum universe are equivalent to that of a 5D vacuum one yielding material from geometry \cite{Wesson1992}. A number of exact solutions have been found for 5D cosmologies, see, for example~\cite{Halpern2002,Wesson1995}.

By applying induced matter theory, a 4D BD theory with an effective matter field and a self-interacting potential, which is called a modified Brans-Dicke theory (MBDT), can be obtained from the vacuum 5D BD field equations \cite{Rasouli2014}.

Vacuum solutions for the anisotropic Bianchi type I model, that is Kasner solutions, have been studied in the context of induced matter theory \cite{Halpern2000}. The corresponding equations in Kasner cosmology lead to a few relations among Kasner parameters and postulate that the energy-momentum tensor and the physical quantities of the theory are obtained in terms of the generalized Kasner parameters.

The energy conditions in a theory of gravity place restrictions on the energy-momentum tensor; Since in this case, it is obtained by Kasner solutions, the energy conditions make limitations on the generalized Kasner parameters. The energy conditions force various linear combinations of the energy density and directional pressure to be non-negative and have been used, for instance, to derive many theorems of classical general relativity such as the singularity theorems, the area increase theorem for black holes, and the positive mass theorem without the need to assume a specific equation of state \cite{Hawking1973,Visser1997}. The energy conditions also have been explored in the context of alternative and modified gravity, such as BD, $f(R)$, $f(R,G)$, etc in Refs.~\cite{Atazadeh2014,Yousaf2018,Capozziello2018,Atazadeh2009,Capozziello2014}. Specifically, we want to focus on defining general energy conditions for MBDT and consider the Bianchi type I model as one that describes our universe.

The energy conditions actually will be specified by the Raychaudhuri equation which are addressing the metric and geodesic structure of the Bianchi type I space-time. Consequently, the marked MBDT energy-momentum tensor has to satisfy some inequalities which one refers to as energy conditions for MBDT.

In section~\ref{sec:2}, we introduce the gravitational field equations for the MBDT of gravity and obtain energy-momentum tensor by its components. In section~\ref{sec:3}, we study energy conditions for an imperfect fluid and properties that are caused in the form of some inequalities which restrict the energy density and pressure contributing to the induced energy-momentum tensor in MBDT. In section~\ref{sec:4}, the state of being acceptable for induced energy-momentum tensor in MBDT will be examined on energy conditions so we can comprehend in what ranges, the satisfaction or violation of energy conditions could happen. And finally, we look over a description of the universe at the present time and compare it with what has been assessed already.

\section{\label{sec:2}Modified Brans-Dicke theory (MBDT)}
The action for the BD theory of gravity in a 5D space-time, without any matter field except the BD scalar field $\varphi$, is written as~\cite{Qiang2005,Sen2001}
\begin{equation}
	S^{[5]}=\int \mathrm{d}^5x \; \sqrt{\vert g^{[5]} \vert}\left[ \varphi R^{[5]} - \frac{\omega}{\varphi} g^{ab} (\nabla\!_a\, \varphi)(\nabla\!_b\, \varphi) \right]
	\;,
	\label{eq:1}
\end{equation}
Here we have taken Einstein gravitational constant in natural units by choosing $8 \pi G = {c^4} = 1$. $R^{[5]}$ is the Ricci scalar associated with the 5D space-time metric $g_{ab}$, the BD parameter $\omega$ is a dimensionless coupling constant, Latin indices run from 0 to 4, and the scalar field which is an equivalent to the inverse of Newtonian constant of gravitation $G$ acts as a gravitational constant; There is also no self-interacting potential considered manually to have a cosmological acceleration. Using the principle of least action, we obtain the field equation of $g_{ab}$ and $\varphi$ derived from~(\ref{eq:1}) respectively as
\begin{eqnarray}
	G_{ab}^{[5]}&=&\frac{\omega}{\varphi^2}\left[{\left( {{\nabla\!_a\,}\varphi} \right)\left( {{\nabla\!_b\,}\varphi} \right) - \frac{1}{2}{g_{ab}}\left( {{\nabla\!_c\,}\varphi} \right)\left( {{\nabla^c}\varphi} \right)} \right]\nonumber\\
	&& + {\frac{1}{\varphi}}\left( {{\nabla\!_a}{\nabla\!_b\,}\varphi - {g_{ab}}{\nabla^c}{\nabla\!_c\,}\varphi} \right)
	\;,
	\label{eq:2}
\end{eqnarray}
and
\begin{equation}
	{\nabla^c}{\nabla\!_c\,}\varphi=0\;.
	\label{eq:3}
\end{equation}
In order to give rise to the present accelerated expansion of the universe, one may add a self-interacting potential manually in the action~(\ref{eq:1}), but instead, we consider the method used in Refs.~\cite{DeLeon2010,Rasouli2011} so that the additional dimension in the BD theory leads to an induced scalar potential in the context of reduced BD theory in 4D. Such a potential which is consisted of geometrical terms can cause the late-time acceleration of the universe.

Taking the dimensional reduction approach, one assumes that the 5D space-time is made up of 4D hypersurfaces $\Sigma_l$ distinguished by certain values for their additional dimension $\Sigma_{l_0}:l_0=$ \textit{constant} and so we divide the metric in 5D into
\begin{eqnarray}
	\mathrm{d}{S^2}&=&{g_{ab}}\left( {{x^c}} \right)\mathrm{d}{x^a}\mathrm{d}{x^b}
	\nonumber\\
	&=&{g_{\alpha \beta }}\left( {{x^\gamma },l} \right)\mathrm{d}{x^\alpha }\mathrm{d}{x^\beta } + \epsilon {h^2}\left( {{x^\gamma },l} \right)\mathrm{d}{l^2}\;,
	\label{eq:4}
\end{eqnarray}
where Greek indices run from 0 to 3, the metric $g_{\alpha \beta}$ is induced on hypersurfaces and $\epsilon$ selects a value between $-1$ and $+1$ relevant to the metric signature convention. each hypersurface is orthogonal to the 5D unit vector along the fifth dimension which is written as
\begin{equation}
	{\hat n}^a = \frac{\delta _{[4]}^a}{h} ~,~~ {n_a}{n^a}=\epsilon
	\;.
	\label{eq:5}
\end{equation}
By assuming that $\hat{n}^a$ is a Killing vector, all components of the metric in 5D space-time would only depend on the cosmic time $g_{ab}(x^c)=g_{ab}(x^0)$.

Employing the dimensional reduction method above at the field equations~(\ref{eq:2}) and (\ref{eq:3}) in 5D, the effective field equations induced on 4D hypersurfaces can be written~\cite{DeLeon2010a}
\begin{eqnarray}
	G_{\alpha \beta}^{[4]} = \frac{1}{\varphi}T_{\alpha \beta}^{\mathrm{(BD)}} + \frac{\omega}{\varphi^2} \left[ {\left( {{{\mathcal D}\!_\alpha}\,\varphi} \right) \left( {{{\mathcal D}\!_\beta}\,\varphi} \right) - \frac{1}{2}{g_{\alpha \beta}} \left( {{{\mathcal D}\!_\gamma}\,\varphi} \right) \left( {{{\mathcal D}^\gamma}\varphi} \right)} \right]
	\nonumber \\
	+ \frac{1}{\varphi} \left( {{{\mathcal D}\!_\alpha}{{\mathcal D}\!_\beta}\,\varphi - {g_{\alpha \beta}}{{\mathcal D}^\gamma}{{\mathcal D}\!_\gamma}\,\varphi} \right) - \frac{1}{2\varphi}{g_{\alpha \beta}}\,V
	\;,
	\label{eq:6}
\end{eqnarray}
and
\begin{equation}
	{{\mathcal D}^\gamma}{{\mathcal D}\!_\gamma}\,\varphi = \frac{1}{2\omega + 3}{T^{\mathrm{(BD)}}} + \frac{1}{2\omega + 3} \left( {\varphi \frac{\mathrm{d}V}{\mathrm{d}\varphi} - 2V} \right)
	\;,
	\label{eq:7}
\end{equation}
where $\mathcal{D}_\alpha$ is the covariant derivative on 4D hypersurfaces. They are exactly similar to the BD field equations in 4D with a self-interacting potential V and an energy-momentum tensor $T_{\alpha \beta}^{\mathrm{(BD)}}$ which are induced on hypersurfaces as the impact of the fifth dimension and consequently give rise to our modified BD theory \cite{Aguilar2008}. The induced energy-momentum tensor and its trace and the induced potential for modified BD theory in~(\ref{eq:6}) and (\ref{eq:7}) are respectively given by
\begin{equation}
	T_{\alpha \beta}^{\mathrm{(BD)}} \equiv \frac{\varphi}{h}{{\mathcal D}\!_\alpha}{{\mathcal D}\!_\beta}\,h + \frac{1}{2}{g_{\alpha \beta}}V
	\;,
	\label{eq:8}
\end{equation}
\begin{equation}
	{T^{\mathrm{(BD)}}} = \frac{\varphi}{h}{{\cal D}^\gamma }{{\cal D}\!_\gamma }\,h + 2V
	\;,
	\label{eq:9}
\end{equation}
and
\begin{equation}
	\frac{\mathrm{d}V}{\mathrm{d}\varphi} \equiv -2\left( {\omega + 1} \right)\frac{\left( {{{\cal D}\!_\gamma}\,h} \right)\left( {{{\cal D}^\gamma }\varphi } \right)}{h\varphi}
	\;.
	\label{eq:10}
\end{equation}
The induced scalar potential $V$ is defined by using the extra dimension and scalar field; The induced energy-momentum tensor in (\ref{eq:8}) is induced geometrically from the fifth dimension of the metric (\ref{eq:4}). It is important to note that only the metric related to 4D hypersurfaces interacts with the matter, since we applied dimensional reduction to the 5D theory before introducing the effective field equation (\ref{eq:6}); The scalar field acts as an intermediary in such a way that it generates a gravitational field along with the matter. On the other hand, the scalar field cannot respond directly to the matter.

Considering the 5D space-time (\ref{eq:4}), we introduce the Bianchi type I universe for the anisotropic model\footnote{
	Wald theorem \cite{Wald1983} implies that if one includes a positive cosmological constant ($\Lambda > 0$) in the Bianchi type I spacetime, consequently, as time passes, the dynamics tends to assume a form similar to de Sitter spacetime. Specifically, each of the scale factors in a Bianchi type I spacetime with a positive cosmological constant becomes, over time, $\mathrm{exp}(\sqrt{\Lambda/3}t)$. Therefore, all of the scale factors become identical and turn the homogeneous and anisotropic spacetime asymptotically into a vacuum de Sitter state.  Such a process is equivalent to the cosmic no-hair theorem \cite{Carroll2018}. In other words, according to the cosmic no-hair theorem, adding a positive cosmological constant to an anisotropic 5D Bianchi type I spacetime leads asymptotically to an isotropic state. For that reason, we did not include a cosmological constant in our model, as that would remove its interesting anisotropic dynamics.}
as
\begin{equation}
	\mathrm{d}{s^2} = - \mathrm{d}{t^2} + {a^2}(t)\,\mathrm{d}{x^2} + {b^2}(t)\,\mathrm{d}{y^2} + {c^2}(t)\,\mathrm{d}{z^2} + {h^2}(t)\,\mathrm{d}{l^2} 
	\;,
	\label{eq:11}
\end{equation}
where $t$ is the cosmic time, $(x,\,y,\,z,\,l)$ are four spatial coordinates and $a(t)$, $b(t)$, $c(t)$, $h(t)$ are different cosmological scale factors in each of the four directions. Then we can derive the effective energy density and directional pressure (where one can define ${\rho}^{\mathrm{(BD)}} \equiv -T_0^{0\, \mathrm{(BD)}}$ and also ${P_i}^{\mathrm{(BD)}} \equiv T_i^{i\,\mathrm{(BD)}}$ by interchanging components among $i=1,\,2,\,3$) from (\ref{eq:8}) respectively
\begin{equation}
	{\rho ^{\mathrm{(BD)}}} \equiv \frac{\varphi \ddot h}{h} - \frac{1}{2}V
	\;,
	\label{eq:12}
\end{equation}
and
\begin{equation}
	P_i^{\mathrm{(BD)}} \equiv - \frac{\varphi \dot h}{h}{H_i} + \frac{1}{2}V ~,~~ i=1,2,3
	\;,
	\label{eq:13}
\end{equation}
where, $H_i$ is the corresponding Hubble parameter in each direction. In order to show how variables in relation~(\ref{eq:11}) and scalar field are related to the cosmic time, we establish the following parameters to make it easier for upcoming calculations.
\begin{equation}
	\left\lbrace
	\begin{array}{l}
		\varphi (t) = {\varphi _0}{t^{{p_0}}}\\
		a(t) \propto {t^{{p_1}}}\\
		b(t) \propto {t^{{p_2}}}\\
		c(t) \propto {t^{{p_3}}}\\
		h(t) = {h_0}{t^{{p_4}}}
	\end{array} \right.
	\;,
	\label{eq:14}
\end{equation}
where the $\varphi_0$ and $h_0$ are constants and the power of $p_a$’s for cosmic time must satisfy all relation obtain from field equations~(\ref{eq:2}) and (\ref{eq:3}); For convenience in our calculations, we set $h_0=1$ without loss of generality. Firstly, we substitute ansatz~(\ref{eq:14}) into equation~(\ref{eq:3}) in a separated form, using coordinates and scalar factors which gives
\begin{equation}
	{p_0}\left( {\sum\limits_{a = 0}^4 {{p_a}} - 1} \right){t^{{p_0} - 2}} = 0
	\;,
	\label{eq:15}
\end{equation}
and this relation produces the important result below
\begin{equation}
	{\sum\limits_{a = 0}^4 {{p_a}}} = 1
	\;.
	\label{eq:16}
\end{equation}
Then we repeat substituting ansatz~(\ref{eq:14}) and expand all of their components in equation (\ref{eq:2}) and finally, with a few manipulations we get to the following constraints among generalized Kasner parameters:
\begin{equation}
	\left( {\omega + 1} \right){{p_0}^2} + \sum\limits_{m = 1}^4 {p_m^2} = 1
	\;,
	\label{eq:17}
\end{equation}
and
\begin{equation}
	\sum\limits_{\mathop {m,n = 1}\limits_{m < n} }^4 {{p_m}{p_n}} = \frac{\left( {\omega + 2} \right)}{2} {p_0^2} - {p_0}
	\;.
	\label{eq:18}
\end{equation}
Now we derive the induced scalar potential $V$ the energy density ${\rho}^{\mathrm{(BD)}}$ and directional pressure ${P_i}^{\mathrm{(BD)}}$ of the induced matter by implying power-law relations (\ref{eq:14}) in equations~(\ref{eq:10}), (\ref{eq:12}), and (\ref{eq:13}) and considering constraints (\ref{eq:16})–(\ref{eq:18}) and setting the constants of integration equal to zero if needed,
\begin{equation}
	V(\varphi) = \left\lbrace
	\begin{array}{ccc}
		{2\left( {\omega + 1} \right)\frac{{p_0^2{p_4}}}{{p_0} - 2}\,{\varphi _0^\frac{2}{p_0}}\,{\varphi^\frac{{p_0}-2}{p_0}}} & \mathrm{for} & {{p_0} \ne 0, 2}
		\medskip \\
		{4{\varphi_0}{p_4}\left( {\omega + 1} \right)\ln \left( \varphi \right)} & \mathrm{for} & {{p_0} = 2}
	\end{array}
	\right.
	\;,
	\label{eq:19}
\end{equation}
\begin{equation}
	{\rho}^{\mathrm{(BD)}} = \left\lbrace
	\begin{array}{ccc}
		{{\varphi _0}{p_4}\left[ {\left( {{p_4} - 1} \right) - \left( {\omega + 1} \right)\frac{p_0^2}{{p_0} - 2}} \right]{t^{{p_0} - 2}}} & \mathrm{for} & {{p_0} \ne 2}
		\medskip \\
		{{\varphi _0}{p_4}\left[ {\left( {{p_4} - 1} \right) - {p_0}\left( {\omega + 1} \right)\ln \left( {{\varphi _0}\,{t^{{p_0}}}} \right)} \right]} & \mathrm{for} & {{p_0} = 2}
	\end{array}
	\right.
	\;,
	\label{eq:20}
\end{equation}
and
\begin{equation}
	{P_i}^{\mathrm{(BD)}} = \left\lbrace
	\begin{array}{ccc}
		{ - {\varphi _0}{p_4}\left[ {{p_i} - \left( {\omega + 1} \right)\frac{p_0^2}{{p_0} - 2}} \right]{t^{{p_0} - 2}}} & \mathrm{for} & {{p_0} \ne 2}
		\medskip \\
		{ - {\varphi _0}{p_4}\left[ {{p_i} - {p_0}\left( {\omega + 1} \right)\ln \left( {{\varphi _0}\,{t^{{p_0}}}} \right)} \right]} & \mathrm{for} & {{p_0} = 2}
	\end{array}
	\right.
	\;.
	\label{eq:21}
\end{equation}
We assume that $\varphi_0 = 1$ for simplicity. For $\omega = -1$, the induced scalar potential in (\ref{eq:19}) vanishes and that value for coupling parameter in BD gravity is predicted as the low-energy limit of the bosonic string theory \cite{Fradkin1985}. In order to have a ghost-free scalar field in (\ref{eq:7}) to prevent making a negative kinetic energy term, we should have $\omega > -3/2$ and the late time accelerated expansion appeals for $-3/2 < \omega < -1$ \cite{DeLeon2010a}. For checking whether the induced energy-momentum tensor in the case of Bianchi type I obeys the conservation law as it does in Einstein theory, one should substitute relations (\ref{eq:20}) and (\ref{eq:21}) for different values of $p_0$ in the relation below for an imperfect fluid
\begin{equation}
	{{\dot \rho}^{\mathrm{(BD)}}} + \sum\limits_{i = 1}^3 {\left( {{\rho ^{\mathrm{(BD)}}} + P_i^{\mathrm{(BD)}}} \right){H_i}} = 0
	\;,
	\label{eq:22}
\end{equation}
and one can prove that the conservation law is satisfied as it is expected.

There is another solution for equation (\ref{eq:15}). Choosing $p_0 = 0$ implies that the scalar field is constant and does not depend on time; The relation (\ref{eq:10}) vanishes and the induced scalar potential $V$ can be zero, without loss of generality. Based on the context we previously referred~\cite{Rasouli2011}, considering the relation~(\ref{eq:16}) leads to generalized Kasner spacetime in 5D which is discussed in~\cite{Rasouli2014}. We ignore $p_0 = 0$ in our study from this point on.

We can now look over different energy conditions using these equations after a quick look at what they are exactly in the general relativity context.

\section{\label{sec:3}Energy conditions}
The energy-momentum tensor in MBDT is physically reasonable when one studies the energy conditions and makes them all satisfied. Using Raychaudhuri’s equation, energy conditions follow from requiring that
\begin{equation}
	{R_{\alpha \beta}}{\upsilon ^\alpha}{\upsilon ^\beta} = \frac{1}{\varphi} \left( {{T_{\alpha \beta}} - \frac{1}{2}{g_{\alpha \beta}}\,T} \right){\upsilon ^\alpha}{\upsilon ^\beta} \ge 0
	\;,
	\label{eq:23}
\end{equation}
where $R_{\alpha \beta}$ is Ricci tensor and $\upsilon ^\alpha$ is a tangent vector to a congruence of time-like geodesics. The term $R_{\alpha \beta}\,{\upsilon ^\alpha }{\upsilon ^\beta} \ge 0$ represents that energy conditions are obtained for the expansion of the universe due to the attractive nature of gravity which can be written in terms of the energy-momentum tensor $T_{\alpha \beta}$ \cite{Atazadeh2014}.

Now we introduce energy conditions and make use of them in modified BD theory by applying relations~(\ref{eq:12}) and (\ref{eq:13}). All kinds of matter must obey these three energy conditions to be accepted as physically reliable ones. So, any theory for a cosmological model which violates these conditions are not trustworthy.

From this point on, we consider a time-like vector $u ^\alpha$ as a four-velocity of an arbitrary observer in space-time which is normalized as $u^\alpha u_\alpha = - 1$ for the signature $\left(-\,+\,+\,+\right)$.
\subsection{\label{sec:3.1}Strong energy condition (SEC)}
For any time-like unit vector $\hat{u} ^\alpha$ at each point of the space-time, there should be a restriction on the energy-momentum tensor set as $T_{\alpha \beta}^\mathrm{(BD)}{{\hat{u}}^\alpha}{{\hat{u}}^\beta} - \frac{1}{2}{T^\mathrm{(BD)}} \ge 0$ which is an inequality about the Ricci tensor $R_{\alpha \beta}\,{\upsilon ^\alpha}{\upsilon ^\beta} \ge 0$ and, further, this inequality can be written in the orthonormal basis,
\begin{eqnarray}
	&&{\rho ^{{\mathrm{(BD)}}}} + P_i^{{\mathrm{(BD)}}} \ge 0 \;, \nonumber \\
	&&{\rho ^{{\mathrm{(BD)}}}} + \sum\nolimits_i {P_i^{\mathrm{(BD)}}} \ge 0 ~,~~ i=1,2,3 \; ,
	\label{eq:24}
\end{eqnarray}
which is the relation we reached for SEC. considering this condition and Raychaudhuri equation \cite{Raychaudhuri1955}, one obtains that gravity has to be attractive.

SEC can be violated if the total energy density $T_{\alpha \beta}^\mathrm{(BD)} {{\hat{u}}^\alpha}{{\hat{u}}^\beta}$ is negative or there should exists a large negative pressure of $T_{\alpha \beta}^\mathrm{(BD)}$ in $T_{\alpha \beta}^\mathrm{(BD)}{{\hat{u}}^\alpha}{{\hat{u}}^\beta} > 0$.

Violation of SEC describes a universe that could have an accelerating cosmic expansion and implies that gravity represents a transition from being attractive to repulsive which is just like the behavior of dark energy for cosmic acceleration. Current observations reveal that the universe has experienced the violation of SEC rather late, somewhere between galaxy formation and the present time.

\subsection{\label{sec:3.2}Weak energy condition (WEC)}
The WEC states that at each event in space-time, we have the local energy density $T_{\alpha \beta}^\mathrm{(BD)}{{\hat{u}}^\alpha}{{\hat{u}}^\beta} \ge 0$ for any time-like unit vector ${{\hat{u}}^\alpha}$. In another word, the energy density for all arbitrary observers is non-negative.

The canonical form of the energy-momentum tensor \cite{Capozziello2018} on the orthonormal basis is
\begin{eqnarray}
	&&{\rho ^{{\mathrm{(BD)}}}} \ge 0 \;, \nonumber \\
	&&{\rho ^{{\mathrm{(BD)}}}} + P_i^{{\mathrm{(BD)}}} \ge 0 ~,~~ i=1,2,3 \; .
	\label{eq:25}
\end{eqnarray}
Violation of WEC leads to violation of second law of thermodynamics for black hole and may yield the cosmic Big Rip. In addition, The singularity theorems by Hawking-Penrose require the validity of both WEC and SEC \cite{Sharif2014}.

\subsection{\label{sec:3.3}Dominant energy condition (DEC)}
Indeed, DEC is similar to WEC with an extra term that the pressure should not exceed the energy density. The DEC can be interpreted as saying that the speed of energy flow of matter for any arbitrary observer can not reach the speed of light. Validity of DEC indicates that for each event in space-time, we have a 4-momentum density vector, $T_{\alpha \beta}^\mathrm{(BD)}{{\hat{u}}^\beta} \ge 0$, which is future pointing and null or time-like.

The canonical form of the energy-momentum tensor \cite{Atazadeh2014} in the orthonormal basis is
\begin{eqnarray}
	&&{\rho ^{{\mathrm{(BD)}}}} \ge 0 \;, \nonumber \\
	&&{\rho ^{{\mathrm{(BD)}}}} \pm P_i^{{\mathrm{(BD)}}} \ge 0 ~,~~ i=1,2,3 \; .
	\label{eq:26}
\end{eqnarray}
Note that DEC implies the WEC. It also seems to be the requirement for all known forms of matter, so a violation of DEC means that the type of matter we are studying is not physically acceptable \cite{Hawking1973}.

\section{\label{sec:4}Validity of induced energy-momentum tensor in MBDT through energy conditions}
Three significant energy conditions are discussed based on a well-known classification of stress-energy tensors introduced in~\cite{Hawking1973} and we sketched a brief interpretation of the satisfaction and violation of these energy conditions for a fluid that has a deviation from being isotropic in spatial directions; it has shown by directional pressure contributing to the energy-momentum tensor of the imperfect fluid that induced from the extra dimension in MBDT.

We note in our study an important analysis by Pimentel, Lora-Clavijo, and González \cite{Pimentel2016} that used novel methods in 4D differential geometry in examining how to define energy conditions for an imperfect fluid. Our methodology is consistent with their analysis, with, in our case, establishing the anisotropic 4D fluid from the 5D geometry based on Wesson’s protocol, and noting that the heat flow in our model is equal to zero.

\subsection{\label{sec:4.1}An attempt to positive energy}
The restrictions defined by inequalities in section~\ref{sec:3} demand the induced energy-momentum of MBDT to be limited in order to declare that energy should be positive so we hereby study the ranges that energy conditions are allowed \cite{Curiel2017}.

Note that, there are other ranges that can satisfy the energy conditions, but induced energy density behaves discretely rough in those ranges. As already mentioned, we assumed that the coupling constant is limited $-3/2~<~\omega~<~-1$. Incidentally, we also separate limited ranges of $p_0$ for simplicity of working with relations (\ref{eq:19}), (\ref{eq:20}), and (\ref{eq:21}); So let us note the limitations below in which relations of energy conditions (\ref{eq:24}), (\ref{eq:25}), and (\ref{eq:26}) would be satisfied when $p_0 \ne 2$:

\medskip \noindent $\bullet$ \quad For SEC to be satisfied in relation (\ref{eq:24}), we have
\begin{eqnarray}
	&&{p_4}\left( {{p_4} - {p_i} + 1} \right) > 0 \;, \nonumber \\
	&&{p_4}\left[ {2\left( {{p_4} - 1} \right) + {p_0} + 2\left( {\omega + 1} \right)\frac{p_0^2}{{p_0} - 2}} \right] > 0 \; .
	\label{eq:27}
\end{eqnarray}
$\bullet$ \quad For WEC to be satisfied in relation (\ref{eq:25}), we have
\begin{eqnarray}
	&&{p_4}\left[ {\left( {{p_4} - 1} \right) - \left( {\omega + 1} \right)\frac{p_0^2}{{p_0} - 2}} \right] > 0 \;, \nonumber \\
	&&{p_4}\left( {{p_4} - {p_i} + 1} \right) > 0 \; .
	\label{eq:28}
\end{eqnarray}
$\bullet$ \quad For DEC to be satisfied in relation (\ref{eq:26}), we have
\begin{eqnarray}
	&&{p_4}\left[ {\left( {{p_4} - 1} \right) - \left( {\omega + 1} \right)\frac{p_0^2}{{p_0} - 2}} \right] > 0 \;, \nonumber \\
	&&{p_4}\left[ {\left( {{p_4} - 1} \right) \pm {p_i} - \left( {1 \pm 1} \right)\left( {\omega + 1} \right)\frac{p_0^2}{{p_0} - 2}} \right] > 0 \; .
	\label{eq:29}
\end{eqnarray}
Inequalities (\ref{eq:27})–(\ref{eq:29}) with taking to account the constraint (\ref{eq:16}) lead us to find the appropriate values for $p_a$’s. the induced energy density $\rho ^{\mathrm{(BD)}}$ in general does not vary continuously and smoothly with variable $p_a$’s as previously mentioned. However, Table~\ref{tab:1} represents those contributing ranges of $p_a$’s in detail so that $\rho ^{\mathrm{(BD)}}$ would remain smooth continuous.
\begin{table}[h]
	\caption{\label{tab:1}
		The regions for $p_i$ and $p_4$ satisfying the energy conditions when we vary $p_0 \ne 0,2$~\footnote{The range $p_0 < 2$ does exclude $p_0=0$.}.}
	\begin{ruledtabular}
		\begin{tabular}{c|cc}
			&\multirow{2}{*}{$p_0 < 2$}&
			\multirow{2}{*}{$p_0 > 2$}\\
			&&\\
			\colrule
			\multirow{2}{*}{SEC} & $p_i < p_4 -1$ & $p_i > p_4 -1$ \\
			& $p_4 > 0$ & $p_4 < 0$ \\[7pt]
			\multirow{2}{*}{WEC} & $p_i > p_4 -1$ & $p_i < p_4 -1$\\
			& $p_4 < 0$ & $p_4 > 0$ \\[7pt]
			\multirow{2}{*}{DEC} & $|p_i| < 1 - p_4$ & $p_i < p_4 -1$\\
			& $p_4 < 0$ & $p_4 > 0$
		\end{tabular}
	\end{ruledtabular}
\end{table}

Then we do a check of energy conditions in the situation that $p_0 = 2$; It must be noticed that both induced energy density $\rho ^{\mathrm{(BD)}}$ and directional pressure ${P_i}^{\mathrm{(BD)}}$ are comprised of a natural logarithm term. So the natural logarithm, $\ln(t)$, for any positive value of $t$ less than $1$ is a negative number and for any positive value of $t$ greater than $1$ is a positive number. Therefore, we investigate the appropriate ranges of $p_a$’s parameters to satisfy energy conditions when $p_0 = 2$ in a way in which time is split into two periods $0 \le t < 1$ and $t \ge 1$. That is:

\medskip \noindent $\bullet$ \quad For SEC to be satisfied in relation (\ref{eq:24}), we have
\begin{eqnarray}
	&&{p_4}\left( {{p_4} - {p_i} + 1} \right) \ge 0 \;, \nonumber \\
	&&{p_4}\left[ {\left( {{p_4} - 1} \right) - 8\left( {\omega + 1} \right) \ln(t)} \right] \ge 0 \; .
	\label{eq:30}
\end{eqnarray}
$\bullet$ \quad For WEC to be satisfied in relation (\ref{eq:25}), we have
\begin{eqnarray}
	&&{p_4}\left[ {\left( {{p_4} - 1} \right) - 4\left( {\omega + 1} \right) \ln(t)} \right] \ge 0 \;, \nonumber \\
	&&{p_4}\left( {{p_4} - {p_i} + 1} \right) \ge 0 \; .
	\label{eq:31}
\end{eqnarray}
$\bullet$ \quad For DEC to be satisfied in relation (\ref{eq:26}), we have
\begin{eqnarray}
	&&{p_4}\left[ {\left( {{p_4} - 1} \right) - 4\left( {\omega + 1} \right) \ln(t)} \right] \ge 0 \;, \nonumber \\
	&&{p_4}\left[ {\left( {{p_4} - 1} \right) \pm {p_i} - 4\left( {1 \pm 1} \right)\left( {\omega + 1} \right)\ln(t)} \right] \ge 0 \; .
	\label{eq:32}
\end{eqnarray}
Relations (\ref{eq:30})–(\ref{eq:32}) satisfy energy conditions for two divided periods which are summarized in Table~\ref{tab:2}.
\begin{table}[h]
	\caption{\label{tab:2}
		The regions for $p_i$ and $p_4$ satisfying the energy conditions when $p_0 = 2$.}
	\begin{ruledtabular}
		\begin{tabular}{c|cc}
			&\multirow{2}{*}{$0 \le t < 1$}&
			\multirow{2}{*}{$t \ge 1$}\\
			&&\\
			\colrule
			\multirow{2}{*}{SEC} & $p_i \ge p_4 -1$ & $p_i \le p_4 -1$\\
			& $p_4 \le 0$ & $p_4 \ge 0$ \\[7pt]
			\multirow{2}{*}{WEC} & $p_i \ge p_4 -1$ & $p_i \le p_4 -1$\\
			& $p_4 \le 0$ & $p_4 \ge 0$ \\[7pt]
			\multirow{2}{*}{DEC} & $|p_i| \le 1 - p_4$ & \multirow{2}{*}{N/A}\\
			& $p_4 \le 0$&
		\end{tabular}
	\end{ruledtabular}
\end{table}

It is seen that energy conditions are sensitive about the time when we set $p_0 = 2$ because of the natural logarithm function. There is no satisfying energy condition with arbitrary parameters at all times for this very case. Hence it is recommended in Ref.~\cite{DeLeon2010} that setting $p_4 = 0$ when $p_0 = 2$ is the only option for satisfying energy conditions for all times, but it implies that $T_{\alpha \beta}^\mathrm{(BD)} = 0$. Therefore, the case $p_0 = 2$ will not yield to any prominent outcome in this regime. 

From the other point of view, the exact value for all five $p_a$’s can not be explicitly obtained for the case $p_0 \ne 0,2$ because the constraint (\ref{eq:16}) and inequalities (\ref{eq:27})–(\ref{eq:32}) make it complex, so we use specific values for some cosmological parameters and introduce them in terms of what we have obtained through studying energy conditions in the following subsection.

\subsection{\label{sec:4.2}Depict the energy conditions derived from MBDT}
As it is worth recalling the importance of MBDT by field equations on any 4D hypersurface, we should interpret the properties of matter obtained from the extra dimension. In order to complete what we have discussed already, we consider specific values for parameters to describe a hypothetical universe and check its characteristics. The inequalities in Table~\ref{tab:1} and Table~\ref{tab:2} that are imposed by the energy conditions help us to reach a physical context for the MBDT with restricted parameters.

The recent detection of gravitational waves from a binary neutron star system GW170817 \cite{Abbott2017} heralded the dawn of the age of multi-messenger astronomy, in which electromagnetic signals can be matched with gravitational waves from the same source. Continued detection of such events could help rule out certain classes of alternative theories, including particular categories of scalar-tensor models—or at least restricting their range of parameters \cite{Ezquiaga2017}. We take note of that in our MBDT model, given that we have replaced the gravitational constant of GR with a changing gravitational field, and it could be parameterized by an inverse relationship with cosmic time based on the equivalency to the Newtonian gravitational constant as specifically described in Refs.~\cite{Vijaykumar2021,Verma2021}. As data from future multi-messenger events such as coincidental gravitational waves and electromagnetic signals from neutron star collisions, accumulates, we will hone our MBDT model’s parameters to accommodate any restrictions introduced.

For giving physical details, one usually defines the spatial volume $\bar{V}$, average scale factor $\alpha$, and the mean Hubble parameter $H$; they also lead us to define a few physical quantities that are important in observational cosmology \cite{Sharif2014,Heydari-Fard2007,Sharif2010,Sharif2012} such as the deceleration parameter $q$ and the expansion scalar $\Theta$ turn out to be
\begin{equation}
	\bar{V} \equiv abc \propto {t^{{p_1} + {p_2} + {p_3}}} \;, 
	\label{eq:33}
\end{equation}
\begin{equation}
	\alpha \equiv \sqrt[3]{\bar{V}} \propto t^{\frac{{p_1} + {p_2} + {p_3}}{3}} \;,
	\label{eq:34}
\end{equation}
and
\begin{equation}
	H \equiv \frac{1}{3} \sum\limits_{i = 1}^3 {H_i} = \frac{1}{3} \frac{\dot{\bar{V}}}{\bar{V}} = \frac{\dot{\alpha}}{\alpha} = \frac{{p_1} + {p_2} + {p_3}}{3t} \;,
	\label{eq:35}
\end{equation}
hence, by relation (\ref{eq:16}) one obtains
\begin{equation}
	H = \frac{{p_1} + {p_2} + {p_3}}{3t} = \frac{1 - {p_0} - {p_4}}{3t} \;,
	\label{eq:36}
\end{equation}
and physical quantities are
\begin{equation}
	q = \frac{\mathrm{d}}{\mathrm{d} t}\left( \frac{1}{H} \right) - 1 = \frac{2 + {p_0} + {p_4}}{1 - {p_0} - {p_4}} \;,
	\label{eq:37}
\end{equation}
and
\begin{equation}
	\Theta = 3H = \frac{1 - {p_0} - {p_4}}{t} \;,
	\label{eq:38}
\end{equation}
where we used relations (\ref{eq:16})–(\ref{eq:18}). Deceleration parameter (\ref{eq:37}) points out that $q \ne - 1$. It is intended that the mean Hubble parameter must depend on cosmic time and also deceleration parameter can be considered as a constant.

The equation (\ref{eq:37}) for deceleration parameter indicates that an accelerated case of universe demands for ${p_0} + {p_4} < - 2$ and ${p_0} + {p_4} > 1$ which are equivalent to $- 1 < q < 0$ and $q < - 1$ respectively. Additionally, the expansion scalar $\Theta$ equation (\ref{eq:38}) reveals that the universe is expanding when ${p_0} + {p_4} < 1$ and the universe contracts for ${p_0} + {p_4} > 1$. For an static universe, one must set ${p_0} + {p_4} = 1$, i.e. the mean Hubble parameter does not depend on cosmic time as mentioned, and it is constant. Consequently, the accelerated expansion of the universe corresponds to ${p_0} + {p_4} < - 2$. If we design a model in accordance with the present epoch of the universe, we expect to have a dark energy-dominated universe. In such case, as introduced in Ref.~\cite{deAraujo2005} we take the vacuum energy, or cosmological constant, as a particular case of dark energy in XCDM context in which the universe expands forever as a power law\footnote{
	One could consider it as being an inflationary phase which follows the power law at the present time.}
in the mean scale factor in equation (\ref{eq:34}) for ${p_0} + {p_4} < - 2$. This range is compatible with our expectations from an accelerated expanding universe.

The nine years of WMAP data on cosmological parameters combining with data from finer scale CMB measurements, the direct measurement of the Hubble constant, and baryon acoustic oscillation (BAO) data \cite{Bennett2013} provide the value for deceleration parameter $q^{\mathrm{(WMAP)}} \approx - 0.65$ that is equivalent to ${p_0} + {p_4} \approx - 7.55$ with regard to the equation (\ref{eq:37}). Furthermore, the latest results of Planck Collaboration \cite{Aghanim2020} that assumes the standard $\Lambda$CDM model, infer a determination of Hubble constant for present time which predicts a late time matter density parameter ${\Omega_m} \approx 0.315$. Presuming to have spatial flatness (${\Omega_m} + {\Omega_\Lambda} = 1$), one has dark energy density parameter ${\Omega_\Lambda} \approx 0.685$. Subsequently, the accurate estimate of deceleration parameter obtain from the Planck data is
\begin{equation}
	q^{\mathrm{(Planck)}} = \frac{1}{2} {\Omega_m} + \frac{1 + 3w}{2} {\Omega_\Lambda} \approx - 0.558 \;,
	\nonumber
	\label{eq:39}
\end{equation}
where $w \approx - 1.03$ is the equation of state parameter of dark energy. The deceleration parameter provided by Planck Collaboration $q^{\mathrm{(Planck)}}$ which is in accord with Kasner parameters of equation (\ref{eq:37}) implies that ${p_0} + {p_4} \approx - 5.792$. So both results related to Planck Collaboration and WMAP are in agreement with our limit for an accelerating expanding universe, i.e. $- 1 < q < 0$ and ${p_0} + {p_4} < - 2$.

The big bang model states that at the origin of universe, in the limit $t \rightarrow 0$, the spatial volume and the expansion scalar in equation (\ref{eq:33}) and (\ref{eq:38}) tend to zero and infinity respectively; Thus, the evolution in an extremely small volume of the universe happens with infinite expansion. When $t \rightarrow 0$, the physical quantities $V$, $\rho^{\mathrm{(BD)}}$, and ${P_i}^{\mathrm{(BD)}}$ in relations (\ref{eq:19}), (\ref{eq:20}), and (\ref{eq:21}) must have infinite values in order that the universe be in accord with the big bang model; It can be possible just by restricting ${p_0} < 2$ in the mentioned relations; According to this fact and using the data in Table~\ref{tab:1} for satisfaction of energy conditions, we reach to a limit i.e. ${p_4} < 0$ which implies that the extra dimension shrinks as the time passes and also, the spatial volume increases and the expansion scalar decreases; The physical quantities also tend to zero. Early infinite expansion ends to a negligible one which demands for slowing down of the expansion rate by time.

\begin{figure*}
	\centering
	\includegraphics[width=0.41\linewidth]{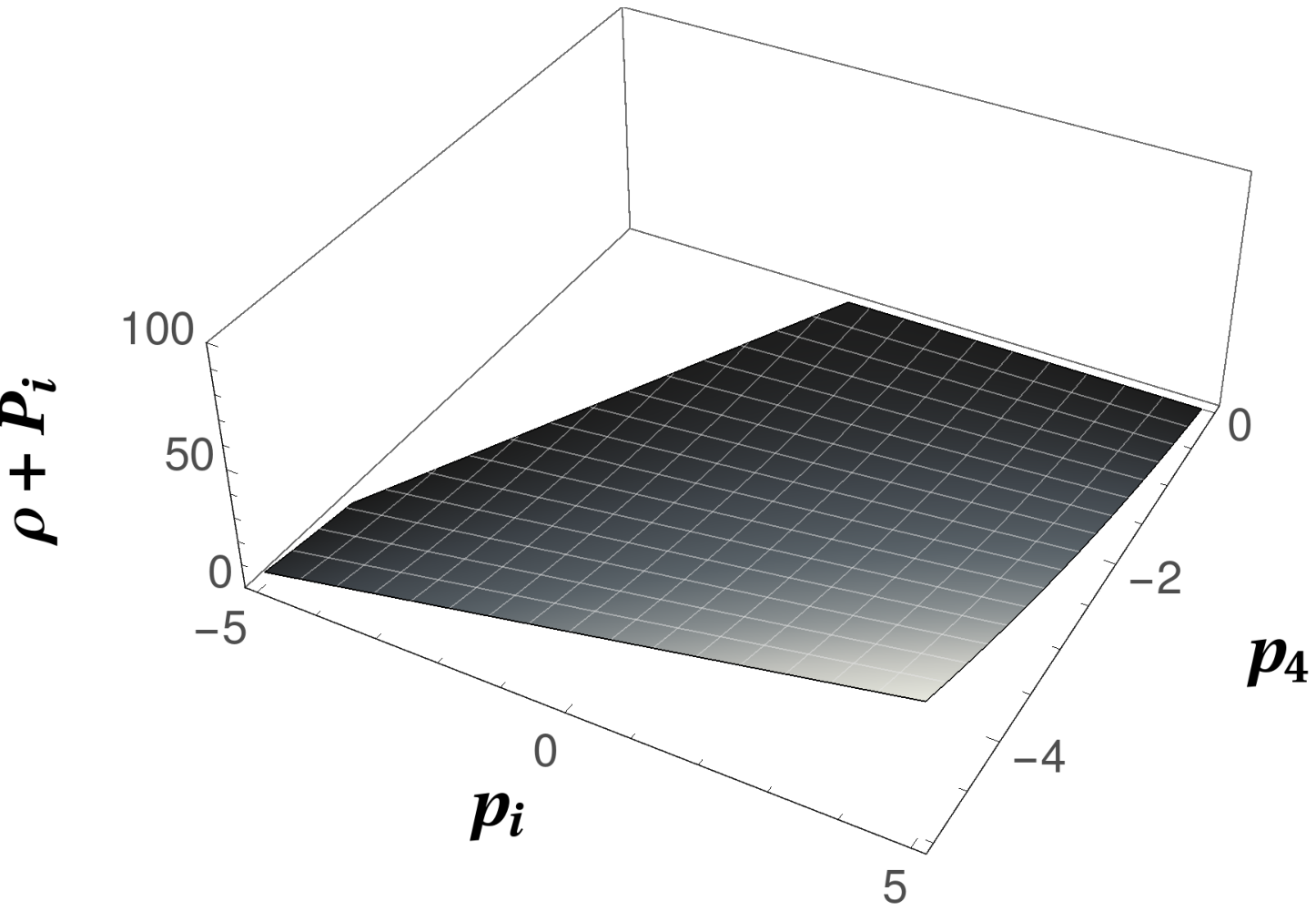}
	~~~~~~~~~~
	\includegraphics[width=0.41\linewidth]{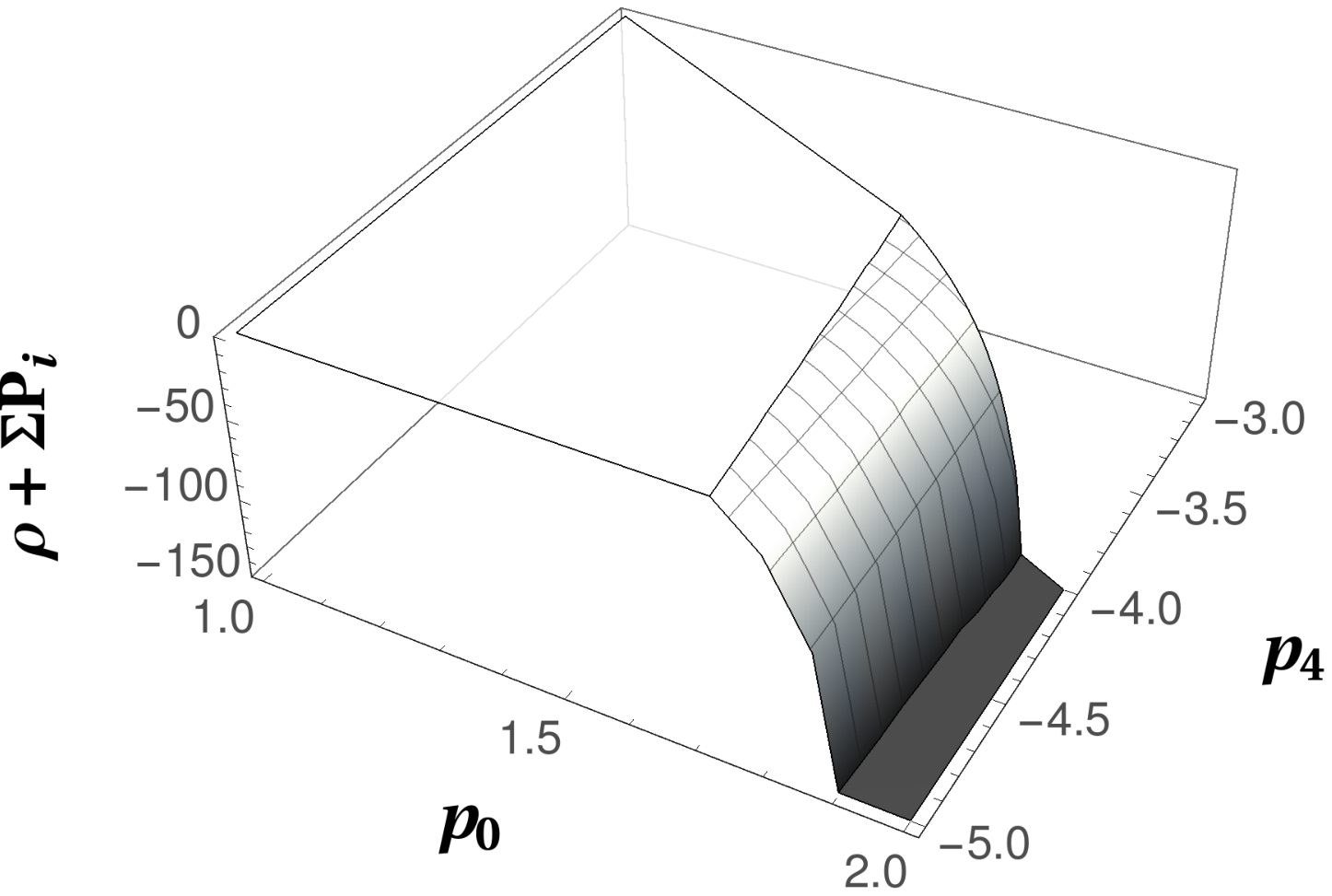}
	\caption{Plots of the strong energy condition for ${p_0} < 2$ and ${p_4} < 0$. The left plot corresponds to ${\rho ^{{\mathrm{(BD)}}}} + P_i^{{\mathrm{(BD)}}} \ge 0$ with $p_i \ge p_4 -1$; The right plot corresponds to ${\rho ^{{\mathrm{(BD)}}}} + \sum\nolimits_i {P_i^{\mathrm{(BD)}}} \le 0$ with $\omega = - 1.25$ and ${p_0} + {p_4} < - 2$. The non-negativity requirement of the SEC is violated for the parameter range considered when $p_0 \rightarrow 2$.}
	\label{fig:sec}
\end{figure*}

\begin{figure*}
	\centering
	\includegraphics[width=0.41\linewidth]{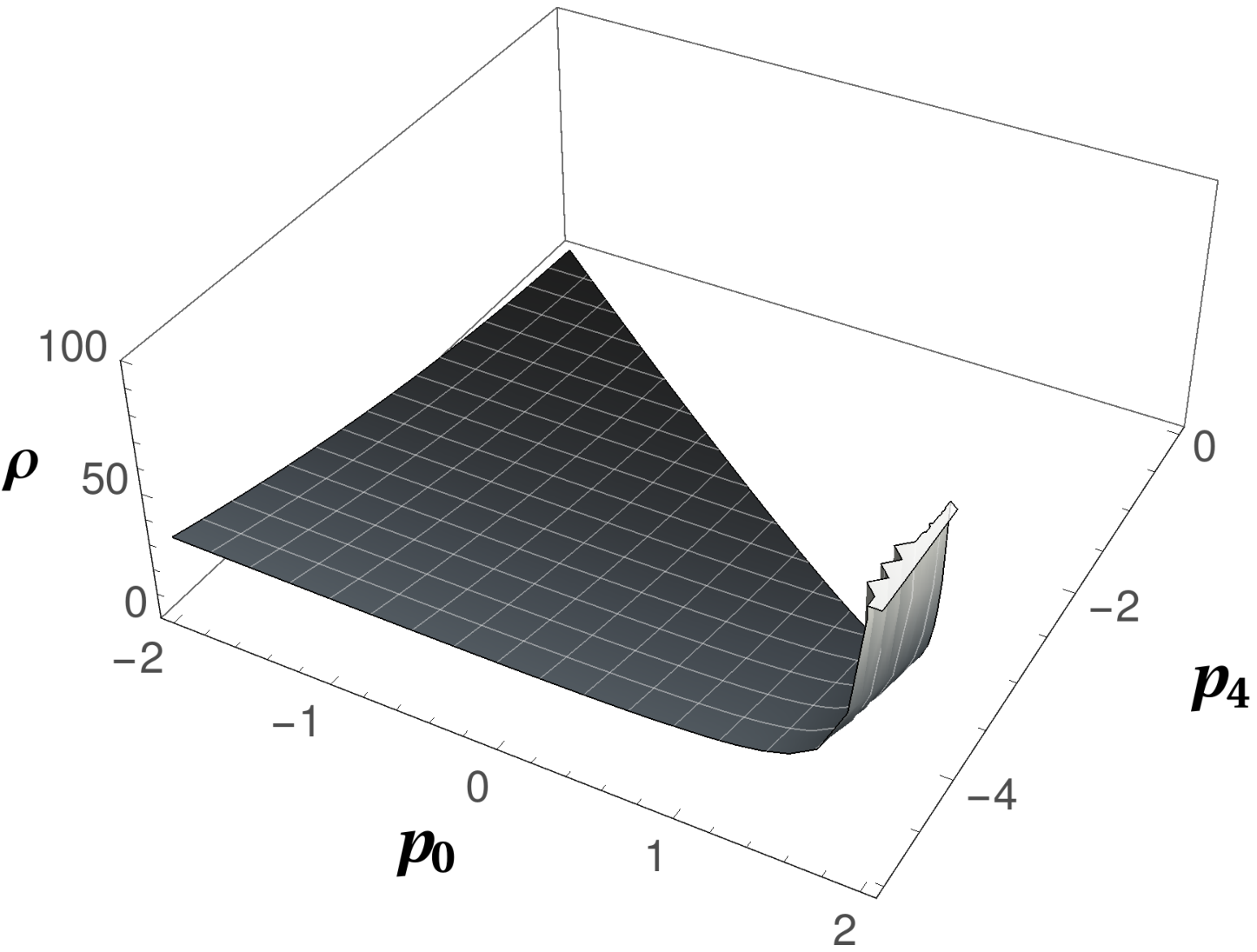}
	~~~~~~~~~~
	\includegraphics[width=0.41\linewidth]{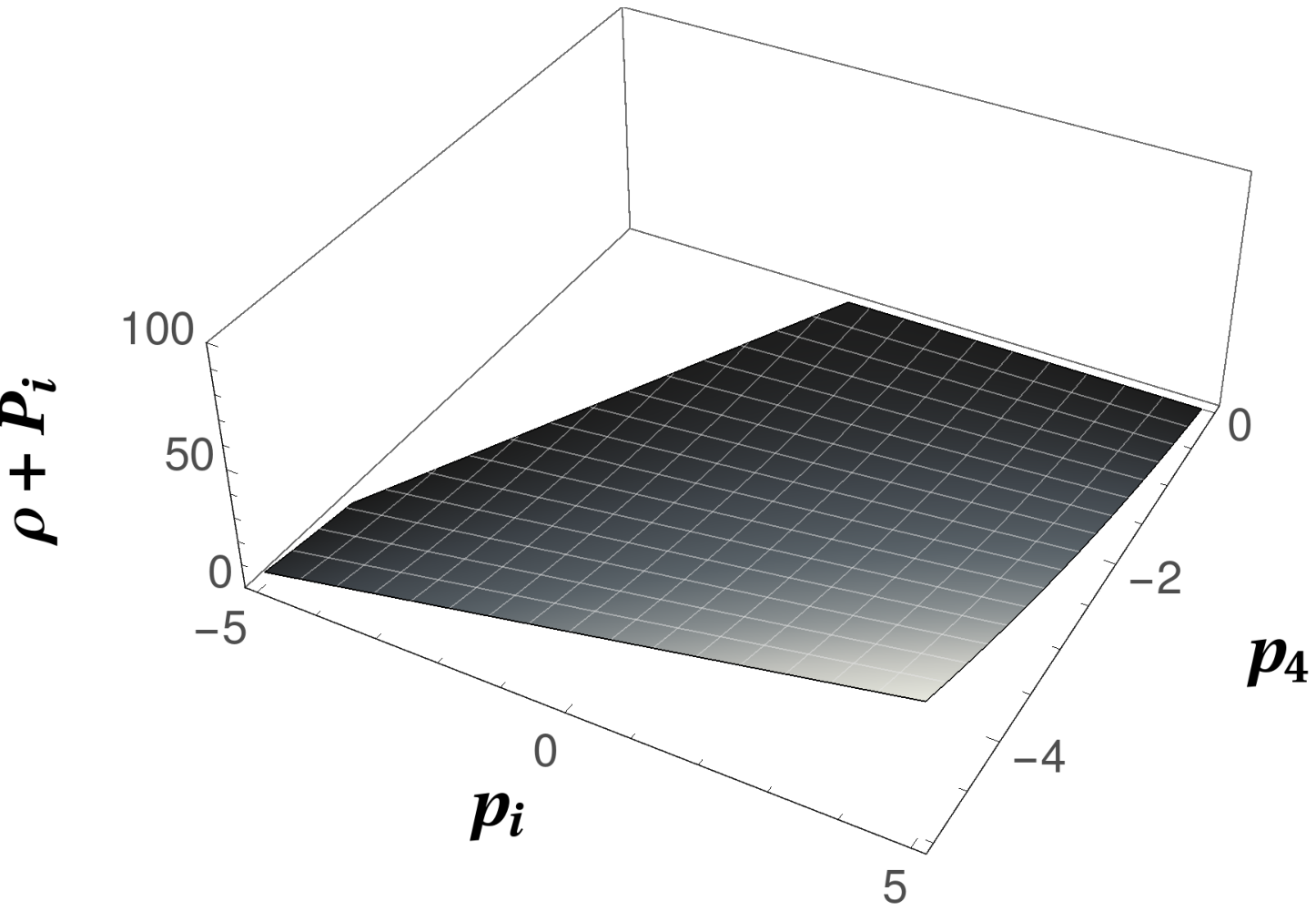}
	\caption{Plots of the weak energy condition for ${p_0} < 2$ and ${p_4} < 0$. The left plot corresponds to ${\rho ^{{\mathrm{(BD)}}}} \ge 0$ with $\omega = - 1.25$ and ${p_0} + {p_4} < - 2$; the right plot corresponds to ${\rho ^{{\mathrm{(BD)}}}} + P_i^{{\mathrm{(BD)}}} \ge 0$ with $p_i \ge p_4 -1$. The non-negativity requirement of the WEC is satisfied for the parameter range considered.}
	\label{fig:wec}
\end{figure*}

\begin{figure*}
	\centering
	\includegraphics[width=0.41\linewidth]{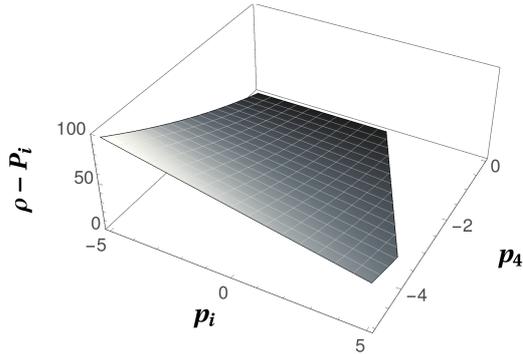}
	\caption{Plots of the dominant energy condition for ${p_0} < 2$ and ${p_4} < 0$ are similar to WECs’ in addition to the one that corresponds to ${\rho ^{{\mathrm{(BD)}}}} - P_i^{{\mathrm{(BD)}}} \ge 0$ shown in this figure. Here, $\omega = - 1.25$, $p_0 = 1.8$, and the limitation $p_i \le \vert p_4 - 1 \vert$ are imposed. The non-negativity requirement of the DEC is completely satisfied for the parameter range considered.}
	\label{fig:dec}
\end{figure*}

Now let us plot the relations (\ref{eq:27})–(\ref{eq:29}) with respect to $p_4$ and $p_i$ or $p_0$ using relevant constraints in Table~\ref{tab:1}, as shown in Figure~\ref{fig:sec}, Figure~\ref{fig:wec}, and Figure~\ref{fig:dec} correspond to SEC, WEC, and DEC respectively. As ${p_0} < 2$ approaches two, before all our power-law relations turn into logarithmic ones ($p_0 \rightarrow 2$ and $p_0 \lesssim 2$), we have a case in which all energy conditions are satisfied except SEC. This agrees with a universe which is undergoing an accelerated expansion; Despite the attractive gravity that is insured by the SEC, repulsive gravity as a result of SEC violation allows for the accelerated cosmic expansion as suggested by the variety of recent observations. It leads to an approach that often accepts the existence of dark energy.

\section{\label{sec:5}Conclusions}
We have investigated the viability of an alternative to BD gravitational theory, namely MBDT by obtaining the general inequalities of energy conditions. One can view the MBDT as a version of induced matter theory in which the BD is the primary theory instead of general relativity.

We have expected that a physically relevant cosmological model must be based on a fluid that does not violate energy conditions. Such a goal is more challenging, but still achievable, with imperfect fluids \cite{Kolassis1988}.

In this paper, we have explored energy conditions for MBDT from the standpoint of the Bianchi type I universe model using an imperfect fluid with different directional pressure—namely, an anisotropic universe model. We have used power-law relations for parameters in MBDT in terms of Kasner parameters. Then, we obtained constraints from energy conditions using those parameters. That allowed us to consider how energy conditions limits are satisfied or violated in the cosmological model, and impose limits accordingly on the parameters related to the MBDT.

In the special case that we investigated, in the present era of our universe, the parameters of MBDT are restricted due to energy conditions. By probing physical values, we demonstrated that the MBDT agrees with our observation and measurements. Restricting the parameters of MBDT by energy conditions led us to an understanding of what characteristics our universe should have. The results matched known present-day dynamics of the universe in terms of its expansion rate, as arising from an infinitesimal volume in the distant past.

\begin{acknowledgments}
The first author (H. Amani) is grateful to Dr. Mehrdad Farhoudi for his guidance and valuable discussions at Shahid Beheshti university in 2019. The second author (P. Halpern) is grateful to the late Paul S. Wesson for his kindness and support.
\end{acknowledgments}

\section*{Data availability}
All data generated or analyzed during this study are included in this published article.

\end{document}